\documentclass[aps,prl,twocolumn,groupedaddress,showpacs]{revtex4-1}
\usepackage[pdftex]{graphicx} 
\usepackage{float} 
\usepackage{amsmath, amsthm, amssymb}
\usepackage{xcolor}

\newcommand{\ds}{\displaystyle}
\newcommand{\beq}{\begin{equation}}
\newcommand{\eeq}{  \end{equation}}
\newcommand{\beqa}{\begin{eqnarray}}
\newcommand{\eeqa}{  \end{eqnarray}}

\newcommand{\grad}{{\boldsymbol \nabla}}

\newcommand{\bem}{\begin{math}}
\newcommand{\eem}{\end{math}}
\newcommand{\rar}{{\rightarrow}}
\newcommand{\Rar}{{\Rightarrow}}
\newcommand{\bfr}{{\bf r}}

\newcommand{\bfv}{{\bf v}}

\newcommand{\bsigma}{{\boldsymbol \sigma}}

\def\strutdepth{\dp\strutbox}
\def\nw#1{\strut\vadjust{\kern-\strutdepth\vtop to0pt{\vss\hbox to\hsize
{\hskip\hsize\hskip5pt$\leftarrow$\hss\strut}}}{\em #1}}
\begin{document}

\title{Pair creation, motion, and annihilation of topological defects in 2D nematics}

\author{Dario Cortese}
\author{Jens Eggers}
\email[]{jens.eggers@bristol.ac.uk}
\author{Tanniemola B. Liverpool}
\email[]{t.liverpool@bristol.ac.uk}
\affiliation{School of Mathematics, University of Bristol,
  Bristol BS8 1TW, United Kingdom}

\begin{abstract}
We present a novel framework for the study of disclinations in
two-dimensional active nematic liquid crystals, and topological defects
in general. The order tensor formalism is used to calculate exact
multi-particle solutions of the linearized static equations inside 
a uniformly aligned state. Topological charge conservation
requires a fixed difference between the number of $q=1/2$ and $q=-1/2$ charges.
Starting from a set of hydrodynamic equations, we derive a low-dimensional
dynamical system for the parameters of the static solutions, which
describes the motion of a half-disclination pair, or of several pairs. Within this formalism, we model defect
production and annihilation, as observed in experiments. Our dynamics also provide an estimate for the critical density
at which production and annihilation rates are balanced.
\end{abstract}

\maketitle
\section{Introduction}
Topological defects are non-trivial configurations of a spatially varying order parameter that are associated with localised singularities~\cite{Mermin1979}.  They are {\em topological} because these singularities can be classified into distinct groups  whose members are  related by a {\em homotopy}~\cite{Milnor_book}. The study of topological defects has a long history:  they have been widely {studied}, for 
example in liquid crystals~\cite{DeGennesProst93,KYHYK02}
optics~\cite{F00,IKSV96,BD07}, and even more recently in biological tissues ~\cite{Saw2017,Kawaguchi2017}. In the last few years, there has been a renewed interest 
from the point of view of topological phase transitions~\cite{Moore2010,Wen2016}. 
Singularities play a crucial role in
determining the structure of many physical problems \cite{EF_book},
and it is therefore a tempting idea to describe the dynamics
of the system by the motion of its singularities. This program
has been followed extensively in describing the motion of
vortices in ideal fluid dynamics \cite{Saff_book}, in the
Ginzburg-Landau equation \cite{Brezis}, or in Bose-Einstein
condensates. 

However, many such approaches are based on {\em dilute} approximations in which the topological defects are  (i) both widely separated 
from each other and (ii)  far from the boundaries~\cite{DeGennesProst93}. The dilute approximation is equivalent to requiring that the deformations induced by each defect to be vanishingly small at the boundaries and in the vicinity of the other defects. If either of these conditions are not satisfied, these problems become much more challenging as defects can no longer be considered independently of each other or the boundaries.

This is because the field surrounding a single defect core is characterized by a singular
phase, which cannot in general be matched to either  to the field at the boundaries (at infinity) or the field near the cores of the other singularities.
In addition, the topology of the space (defined by the Euler characteristic) in which the vector field (e.g. liquid crystalline order) lives imposes constraints on the number and charges of the defects via the Poincar\'e-Hopf theorem~\cite{Milnor_book}.
For example, a consistent treatment requires one  consider
multi-particle states with constraints on the number and charge of the defects, such that the
total charge adds up to the Euler characteristic (zero for a flat plane with no holes). 
Recent experiments on active liquid crystals~\cite{marchetti2013hydrodynamics} provide a motivation to address these longstanding issues as under many conditions, activity leads to `chaotic' states with a proliferation of defects~\cite{Sanchez2012,Sumino2012,Schaller2013,idecamp_orientational_2015} which consequently are not widely separated from each other or boundaries, {\em requiring} one to go beyond the dilute approximation.

In this article we characterize and study the dynamics of topological defects in two-dimensional nematic liquid crystals, though we believe the approach we develop to be 
more generally applicable to other geometric singularities in a variety of physical systems. To be precise,
here we will consider only the lowest energy defects consistent with nematic liquid crystal symmetry, positive and negative 
half-integer defects or disclinations \cite{DeGennesProst93} on a two-dimensional surface. For a plane with no holes, this implies an even number of defects (particles) with equal numbers of positive and negative charges~\cite{Milnor_book}.  Although such particle pairs play
an important role in many famous physics problems, such as
superconductivity (where positive and negative particles form
Cooper pairs), or the Kosterlitz-Thouless transition~\cite{Kosterlitz1973} (which results
from the disassociation of vortex pairs), multi-particle states
are usually not known explicitly. 

However, in the present paper we find explicit expressions for 
many-particle states
of singularities in nematic liquid crystals, so called
disclinations \cite{frank_i._1958}, which have topological
charges of $q=\pm 1/2$. This is particularly exciting since we
are thus able to mathematically describe the creation of a defect-pair itself, where
a pair of oppositely charged particles are formed spontaneously
out of a uniform state. Likewise, we characterize the annihilation of
pairs of defects, where two particles come together to form a
uniform state.
We will describe these singular events for an active suspension of elongated
particles \cite{Sanchez2012,idecamp_orientational_2015} in a
nematic liquid crystal phase.  This is an example of 
active matter driven out of equilibrium by constituents which consume energy, the study of which has emerged recently as an exciting new field in
soft condensed matter \cite{marchetti2013hydrodynamics}. In the experiment, a thin film of microtubules (MT)
is suspended on an oil layer. Molecular motors crosslink MT's and
induce relative sliding, which induces  motion, and
pumps energy into the fluid layer. 

Without activity, the fluid is at rest, and the system relaxes to
a uniformly ordered nematic state, in which all particles are oriented
in the same direction. However,
activity induces a highly non-uniform state, and in particular leads
to the creation of a ``gas'' of defects or disclinations. 
The random arrangement of defects is due to constant pair-creation and
annihilation events.
%
There have been
a number of successful large-scale numerical simulations of this
system~\cite{giomi_banding_2012,Thampi2013,wensink_meso-scale_2012,hemingway_correlation_2016},
based on a standard continuum model of an active fluid
\cite{marchetti2013hydrodynamics}. This will serve as 
a guide for our 
theoretical calculations. 

Previous theoretical attempts at the problem~\cite{pismen_dynamics_2013,giomi_defect_2014,Shi2013} were all based
on the hypothetical dynamics of a {\it single} defect~\cite{P88,RK91}. This requires ad-hoc assumptions on the
form of the far field, and necessarily introduces a
dependence on some length scale, which serves to remove singularities.
It is unknown how to identify this length scale uniquely, based on the
equations of motion. Our aim here then is therefore to formulate a dynamics
for defects based on first principles, relying on the equations
of motion only. 

\section{Statics: multi-defect states}
Let us begin with a description of the equilibrium states of a uniaxial 
nematic crystal, described by its director,
${\bf n} = (\cos\theta,\sin\theta)$, for which the Frank-Oseen free energy
is~\cite{DeGennesProst93}
\beq
F_{\tiny FO} = \frac{K}{2}\int \left|\left|\nabla\mathbf{n}\right|\right|^{2}d{\bf r} =
\frac{K}{2}\int \left|\nabla\theta\right|^{2}d{\bf r} \; .
\label{F}
\eeq
For simplicity, we have used the one-constant approximation
$K\equiv K_1=K_2=K_3$. 
It is crucial
to note that in a nematic crystal, ${\bf n}$ is an axial vector, for which ${\bf n} \equiv -{\bf n}$.  
Similarly, the orientation angle $\theta$
is defined only up to multiples of $\pi$.
Points of stationary variation $\delta F_{FO}/\delta \theta = 0$ define
equilibrium states, solutions of Laplace's equation
\beq
\triangle \theta = 0 \; , 
\label{stat}
\eeq
where $\triangle \equiv {\nabla}^2$.
However, equation \eqref{stat} does not mean that equilibrium states are
defined by a simple linear equation; rather, nonlinearities arise because of the equivalence
$\theta \equiv \theta \pm\pi$. 

\begin{figure}
\includegraphics[width=0.95\columnwidth]{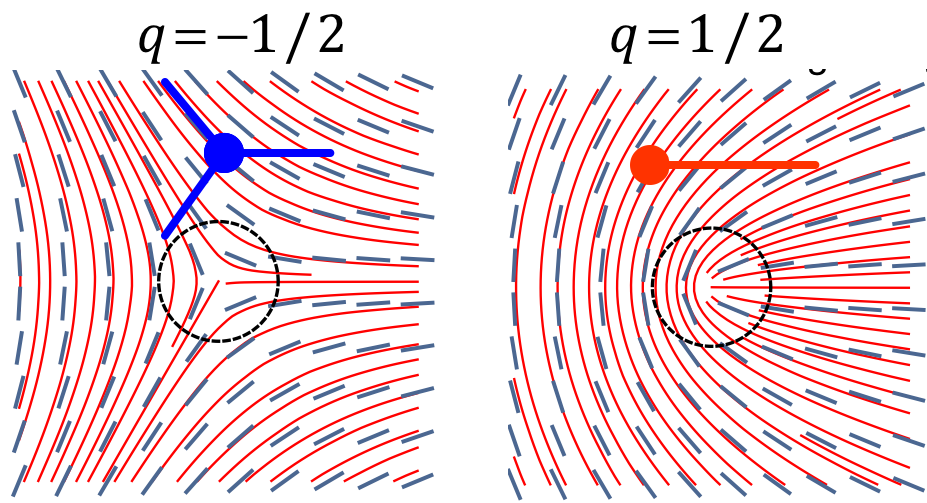}
\caption{The disclination, equation \eqref{sing} for $m=\pm 1$, with charge
  $q=\pm 1/2$. 
\label{fig:pm}}
\end{figure}
It was noted by Oseen \cite{oseen_theory_1933,frank_i._1958}, that equation (\ref{stat})
admits solutions corresponding to the two-dimensional singularities
\beq
\theta_d^{(m/2)}({\bf r}) = \frac{m}{2}\phi \; ,
\label{sing}
\eeq
where $m$ is an integer and ${\bf r} = r(\cos\phi,\sin\phi)$ is
the position vector. The two lowest order disclinations $m=\pm 1$
are shown in Fig.~\ref{fig:pm}. Half-integer values of the prefactor
are allowed in equation (\ref{sing}), since $\theta = \pm \pi$ is equivalent to
$\theta = 0$, so that the director returns to is original state after a
full rotation. 

Inserting equation \eqref{sing} back into equation \eqref{F}, one finds the
free energy of a single defect $F^{(q)} = \pi K q^2 \ln(L/a)$.
To make the result finite,
we had to introduce a small scale core size $a$ and a large-scale
cutoff $L$. Both scales will be described self-consistently by the
theory we are about to develop. However, it does follow from this
simple estimate, that in a two-dimensional system the excitations most 
likely to occur are the two non-trivial lowest energy states $m=\pm 1$. 

The topological character of a defect is defined by its topological
charge
\bem \ds
q = \frac{1}{2\pi}\oint_{\cal C} d\theta =
\frac{1}{2\pi}\int_{0}^{2\pi} \frac{d\theta}{d\phi}d\phi \; , \; 
\label{charge}
\eem
where ${\cal C}$ is any closed loop around the defect. Clearly, 
for the singular solution, equation \eqref{sing} the result is the charge
$q = m/2$, which can take half-integer values. 
For these half-integer defects, however, there is associated to each defect an attached unbounded singular line at which $\theta$ (equivalently $\bf n$) jumps $\pm \pi$ (the fact that ${\bf n} \equiv -{\bf n}$ means that the singular line is an artefact of the parametrization). This highlights the fact that ${\bf n} ({\bf r})$ is insufficient to describe the singularity completely. 

In order to rectify this problem, we use an expression for the free energy, due to de Gennes
\cite{DeGennesProst93}, which includes the additional physics necessary to
describe the structure of the core of a defect near its center and removes the artificial singular line.
The key is to instead of ${\bf n}$, use as order parameter the symmetric, traceless matrix
\beq
\mathbf{Q}({\bf r})=\left(\begin{array}{cc}
Q_{1} & Q_{2}\\
Q_{2} & -Q_{1}
\end{array}\right)=Q_{0}\left(\begin{array}{cc}
2n_{x}^{2}-1 & 2n_{x}n_{y}\\
2n_{x}n_{y} & 1 - 2n_{x}^{2}
\end{array}\right),
\label{Q}
\eeq
which can be expressed in terms of the director ${\bf n}({\bf r})$
and the degree of alignment $Q_0({\bf r})$. In particular, the symmetry
of ${\bf n}$ is now built into the description.
In order to guarantee a smooth solution at the core, we use 
the Landau-de Gennes free energy 
\beq
F_{LdG}  = \int \left( -\frac{A}{2}\left|\mathbf{Q}\right|^{2}+
\frac{B}{4}\left|\mathbf{Q}\right|^{4}+
\frac{K}{2}\left|\nabla\mathbf{Q}\right|^{2} \right) d{\bf r} \; , 
\label{Free_energy_Q}
\eeq
which allows the amount of nematic ordering to vary.

Furthemore, we note that there is no way a single defect can be placed in a neutral environment
(for example a constant director ${\bf n} = {\bf e}_x$) without $\theta$
encountering a singularity. Embedding defects into a system with a uniform director requires that the total charge vanishes,
which means there must be an equal number of positive and negative
half-charges. Thus in any attempt to 
construct singular solutions which decay to a uniform director field at
infinity, one must automatically contemplate many-particle solutions,
which incorporate charge neutrality. 

The elementary disclinations $q=\pm 1/2$ now
have the local form,  
\bem
\mathbf{Q}= Q_{0}(\bfr)\left(\begin{array}{cc}
\cos\phi & \pm\sin\phi\\
\pm\sin\phi & -\cos\phi
\end{array}\right), \quad q = \pm\frac{1}{2} \; .
\label{nQtheta}
\eem

For ${\bf Q}$ to be smooth near the origin, $Q_0(r)$ must
go to zero for $r\rightarrow 0$, consistent with its interpretation as
a measure of local order: at the center of defect, ${\bf n}$ points in
all directions, so there is no order. As a result, zeroes of
\(
Q_0({\bf r}) = \sqrt{Q_1^2({\bf r}) + Q_2^2({\bf r})},
\)
which are places where $Q_1({\bf r})$ and $Q_2({\bf r})$ vanish
simultaneously, are most conveniently used to find the exact position
of a disclination. 
In the following we will now embed the defects into an environment with a uniform director field.
From a balance of the first two terms of equation (\ref{Free_energy_Q}), one finds a uniform solution
(so that the gradient term disappears) of the form
\bem
Q_1 = \overline{Q}_0\cos\xi, \quad
Q_2 = \overline{Q}_0\sin\xi \; , \;  
\label{uniform}
\eem
where $\xi$ is the (constant) orientation angle and 
$\overline{Q}_0 = \sqrt{2 A/B}$.

Once more, equilibrium states  are found from the
vanishing variation of free energy, ${\bf H} =  - {\delta F_{LdG} / \delta {\bf Q} }$ which leads to the
pair of nonlinear equations
\beq
{\bf H} =0 \;  \Rar \; K\triangle Q_{1,2}+\left[A-2B\left(Q_{1}^{2}+Q_{2}^{2}\right)\right]Q_{1,2}=0 \; .
\label{STATIC_nonlin}
\eeq
It makes explicit all the nonlinearities contained implicitly in
equation \eqref{stat}, and contains additional physics to describe disclinations
using smoothly varying fields $Q_1,Q_2$. 
We are interested in solving equation \eqref{STATIC_nonlin} such that they locally
describe a $q = \pm\frac{1}{2}$ disclination 
yet have a uniform orientation
; without loss of generality we take $\xi=0$, i.e. the nematic is oriented along the $x$-axis.

We linearize
equation \eqref{STATIC_nonlin} around the uniform state, which is given by
$\overline{Q}_1 = \overline{Q}_0 = \sqrt{2A/B}$ and 
$\overline{Q}_2 = 0$:
\(
Q_1 = \overline{Q}_1 + \delta Q_1, \quad
Q_2 = \delta Q_2 \; . \;  
\)
Thus the linear equations become 
\beq
\triangle\delta Q_1-\kappa^2\delta Q_1  = 0, \quad
\triangle \delta Q_2  = 0, 
\label{STATIC_lin}
\eeq
where $\kappa=\ell_{Q}^{-1}=\left(2A/K\right)^{1/2}$ is
the inverse elastic length scale. This length scale also sets the
size of a defect. Linearization of the $\bf Q$ equation makes this problem analytically tractable by assuming variations in $Q_0$ are small, {\em but} retains all the nonlinearities associated with the variation of the director, $\bf n$. It is an improvement on equation \eqref{stat} which assumes $Q_0$ constant.
Once the solution
is found in terms of $Q_1,Q_2$, the orientation can be reconstructed by
inverting the relations 
\bem
Q_1 = Q_0\cos2\theta, \;
Q_2 = Q_0\sin2\theta \; 
\label{orientation}
\eem
to find the orientation angle $\theta (\bfr)$. 

Now we want so solve equation \eqref{STATIC_lin} with boundary condition prescribed at the singularity
at the origin; by construction, $\delta Q_{1,2}$ have to vanish at infinity, giving the other required boundary condition.
The boundary condition at the singularity is specified by a given angular dependence on a circle of radius
$a$ around the origin. The most general ansatz is the Fourier series in $\phi$, $Q_{\alpha}(a,\phi)$ :
\beq
Q_{\alpha}(a,\phi) =
\overline{E}_{\alpha} + \sum_{n=1}^{\infty}\left[\overline{D}_\alpha^{(n)}
\cos\left(n\phi+\zeta_{\alpha}^{(n)}\right)\right] \; , 
\label{bc}
\eeq
where $\alpha=\{1,2\}$. It is here that the topological charge of the {\em imposed defect} is fixed, by the lowest non-zero mode $n$ of 
equation \eqref{bc}.
The length $a$ can be interpreted as the core size of the defect,
which is a microscopic scale, set by the particle size. It is
expected to be much smaller than the elastic length $\kappa^{-1}$,
over which elastic stresses relax. 
%
%
A solution to equation \eqref{STATIC_lin} for $\delta Q_\alpha (r,\phi)$, $\alpha \in \{1,2\}$
is a superposition of Fourier modes of the form \cite{LS03}
\beq
\delta Q_\alpha = \sum_{n}h_{\alpha, n}(r)\left(A\cos n\phi + B\sin n\phi\right) \; . 
\label{eq:fourier}\eeq
Then $h_{1,n}(r)$ are solutions of a modified Bessel equation \cite{GR14}, the
solutions which decay at infinity are $K_n(\kappa r)=\int_0^\infty dt \cosh (n t) e^{- \kappa r \cosh t}$. This describes
the solution for $r>a$, which is the only part of physical interest.
The function $h_{2,n}(r) \sim r^{p(n)}$ is a power law solution of Laplace equation, with $p>0$ for $r< a$ and $p<0$ for $r>a$. 
%
%

We demonstrate below that only the constant and $n=1$ terms of the Fourier series for the boundary conditions, equation \eqref{bc} are required to obtain half-integer disclinations 
and that the free parameters in equations  \eqref{bc} , \eqref{eq:fourier} determine the number, locations and orientations of the defects.
Hence restricting our analysis first to only the constant (zero-mode) and the $n=1$ mode  (easily generalized to higher modes), we require
\(
Q_1 (a,\phi) = \overline{E}_1 + \overline{D}_1\cos(\phi+\zeta_1)\; , \;  
Q_2 (a,\phi) = \overline{E}_2 + \overline{D}_2\sin(\phi+\zeta_2) \; 
\)
on $r = a$. 
The constant $\overline E_2$ ($n=0$ term for $Q_2$) provides both essential information about the defect topology and a technical difficulty, as it does not correspond to a single term of a sine-Fourier series.  In fact it can only be addressed by using an infinite number of terms of the series. To deal with it, we represent it as a sum of
Fourier modes, noting the series for a square pulse between 
$\phi = -\pi$ and $\phi = \pi$ is :
\beq
\overline{E}_2 = \frac{4\overline{E}_2}{\pi}
\sum_{n=0}^{\infty}\frac{(-1)^n}{2n+1}\cos\frac{(2n+1)\phi}{2}. 
\label{pulse}
\eeq
Thus the $n=0$ mode contribution to $\delta Q_2(\bfr)$ can be written as a sum of 
powers $(a/r)^{n+1/2}$, whose coefficients are the terms in the
sum equation \eqref{pulse}.
The resulting expression can be resummed and if we rescale $\delta Q_1$ and $\delta Q_2$ with $\overline{Q}_0$, 
and write $r$ in units of $a$ (such that $r=1$ at the microscopic
size of the defect), we obtain
\beqa
\label{Q1_ren_n=0,1}
&& \delta Q_1 = (E_1-1)\frac{K_{0}\left(\Lambda r\right)}
{K_{0}\left(\Lambda \right)} + D_1 \frac{K_1\left(\Lambda r\right)}
{K_1\left(\Lambda\right)}\cos(\phi + \zeta_1), \\
\label{Q2_ren_n=0,1}
&& \delta Q_2 = D_2\frac{\sin(\phi + \zeta_2)}{r}  + 
E_2 f_2(r,\phi).
\eeqa
where
\begin{eqnarray*}
&& f_2(x,\phi) = \frac{2}{\pi}\left[{\rm arccot}
\left(\frac{\sqrt{x}}{\cos\phi/2}
+\tan\frac{\phi}{2}\right) + \right. \\
&& \left.{\rm arccot}\left(\frac{\sqrt{x}}{\cos\phi/2}
-\tan\frac{\phi}{2} \right)\right]. 
\end{eqnarray*}
This is one of the main results of this paper.

\begin{figure}
\includegraphics[width=\columnwidth]{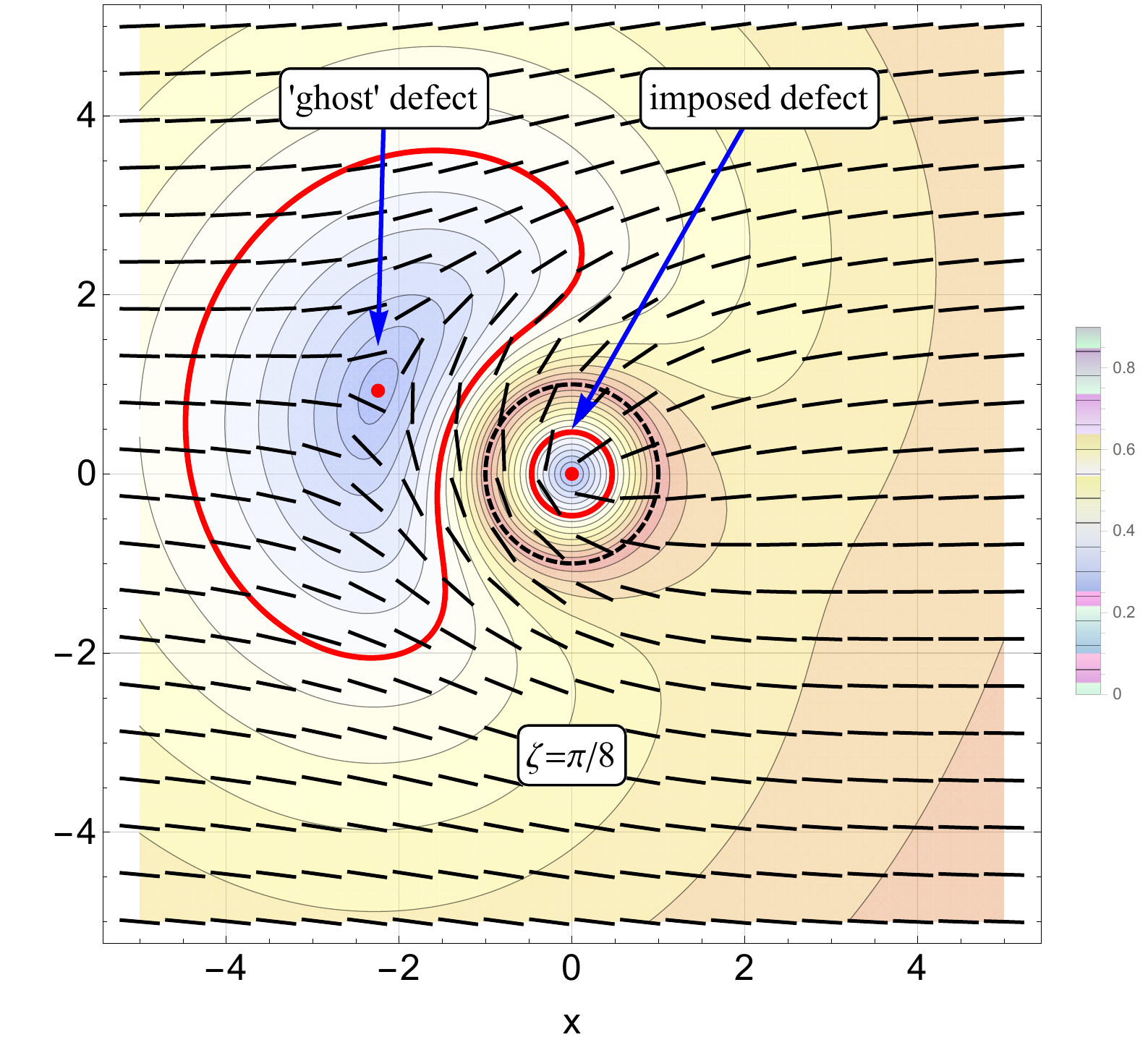}
\includegraphics[width=\columnwidth]{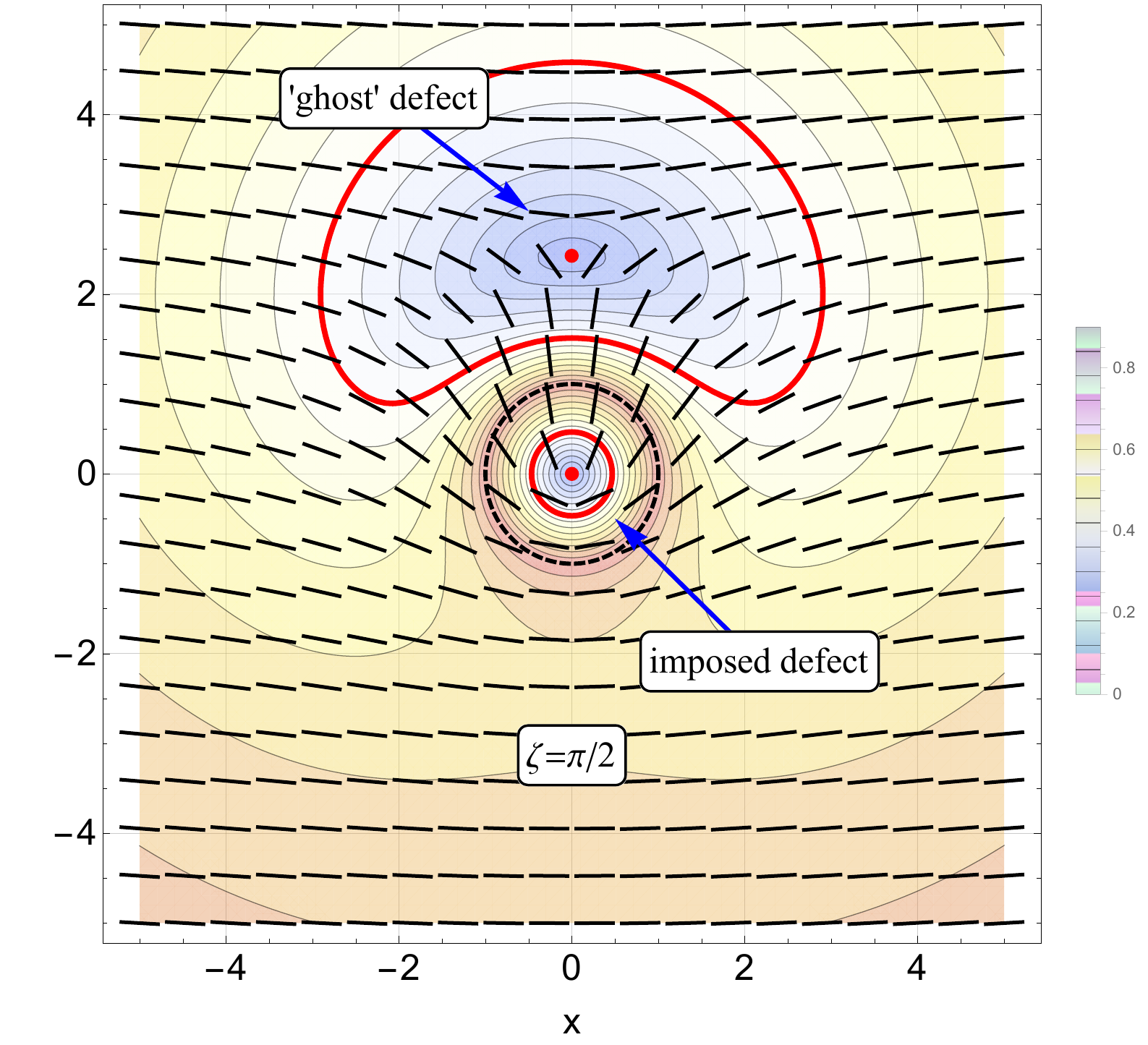}
\caption{Director configuration (black bars) and order tensor magnitude
(contours and colours) for a pair of oppositely charged half-disclinations.  The positive defect was imposed in the solution of
equation \eqref{STATIC_lin}, the negative ``ghost'' emerges to
satisfy the constraint of zero charge. Two typical two-defect configurations, as described
by equations \eqref{Q1_ren_n=0,1},\eqref{Q2_ren_n=0,1}. Here, $D_{1,2}=0.9, E_{1,2}=0,\zeta_{1,2}=\pi/8, \pi/4$}
\label{pair_states}
\end{figure}

%
A couple of examples of  typical director configuration are shown in
Fig.~\ref{pair_states}; 
apart from the imposed $q=1/2$ defect, a second
``ghost'' defect has appeared, whose position and orientation depends on the
parameters chosen. Thus the total charge of the system is zero, and
the director field is uniform far away from the pair. Any solution of
equation \eqref{STATIC_lin} which satisfies uniform boundary conditions must
automatically satisfy charge neutrality. 
%
%

We can thus characterise a
pair of defects in terms of 6 scalar parameters $D_{1/2},E_{1/2}$
and $\zeta_{1/2}$. Two examples are illustrated in Fig.~\ref{pair_states}. 
Choosing $D>0$ or $D<0$, corresponds to charge $q=1/2$ or $q=-1/2$
for the imposed defect, respectively, and thus effectively interchange
the imposed and ghost defects. The angles $\zeta_{1/2}$ control the
orientation of the imposed defect relative to the order in the far field.
The coefficients $E_1$ and $E_2$ can be written as $E_1 = E_0\cos\xi$ and
$E_1 = E_0\sin\xi$, where $E_0$ controls mainly the degree of
anisotropy, whereas $\xi$ is the angle between the two orientations. 
$E_1$ and $D$ are the dominant parameters controlling the distance between defects.

To study the defect dynamics, our strategy will be to obtain a reduced model in terms of equations of motion for
the parameters, and then to use the time dependent parameter values to calculate the time-dependent vortex configurations once
the parameter values have been obtained. 
Equations \eqref{Q1_ren_n=0,1},\eqref{Q2_ren_n=0,1} correspond to states with at most 
two defects. However, by including  more modes, states with arbitrary number of 
defects can be generated (see Appendix).

\section{Defect dynamics: pair creation and annihilation}

We study the temporal dynamics of disclinations using the standard equations of nematodynamics at vanishing Reynolds number in two dimensions augmented to 
include the possibility of additional active stresses~\cite{giomi_banding_2012,marchetti2013hydrodynamics}.
A key component of these 
are the Stokes equations
describing the motion of a viscous nematic  fluid~\cite{giomi_banding_2012,marchetti2013hydrodynamics}.
 They are driven by
the active stress $\bsigma_a = \alpha c_0^2 {\bf Q}$, where $c_0$ is the concentration of active particles, and the elastic stress,
which results from the nematic not being at elastic equilibrium, ${\bf H} \ne 0$, see equation \eqref{STATIC_nonlin}. 
A non-vanishing $\bf H$ indicates an unbalanced
elastic stress, so
$ \bsigma_{el} = -\lambda Q_0 {\bf H} + {\bf QH} - {\bf HQ}$.
If $\alpha<0$ (``pushers''), the active particles are extensile. The
case $\alpha>0$ (``pullers'') corresponds to contractile particles.
The so-called alignment parameter $\lambda$ will be discussed below. 
Both extensile and contractile cases lead generically to instability with increasing $\alpha$, depending on the parameter, $\lambda$.
Thus Stokes' equation for an active incompressible nematic fluid
$\grad\cdot{\bf v} = 0$ becomes
\beq
\eta\triangle\mathbf{v}+\grad\cdot\left[\bsigma_{el} + \bsigma_a \right] =0 \;, 
\label{Stokes}
\eeq
To close the system of equations, we need the equation of motion
for ${\bf Q}$:
\beq
\frac{D \mathbf{Q}}{Dt} = \frac{\mathbf{H}}{\gamma} + 
\lambda Q_0\mathbf{V} 
-\alpha c_{0}(\nabla\cdot\mathbf{Q})\cdot\nabla\mathbf{Q} ,
\label{Q_motion}
\eeq
where $V_{ij} = (\partial_i v_j + \partial_j v_i)/2$ and
$\omega_{ij} = (\partial_i v_j - \partial_j v_i)/2$ are the symmetric
and antisymmetric parts of the velocity gradient tensor $\grad {\bf v}$,
respectively. The corotational derivative 
$D\mathbf{Q}/Dt = \partial_t \mathbf{Q} + \mathbf{v}\cdot\grad\mathbf{Q} +
\boldsymbol{\omega}\mathbf{Q} - \mathbf{Q}\boldsymbol{\omega}$
accounts for the fact that rod-like particles move and rotate with the
fluid.
The first term on the right of equation \eqref{Q} describes the tendency of
the nematic crystal to relax to an elastic equilibrium state, for which
${\bf H} = 0$; this occurs on a time scale $\gamma$. The next term describes
the motion of an elongated particle in shear flow; the dimensionless
parameter $\lambda$ measures the tendency of the particle to align with
the flow \cite{edwards_spontaneous_2009}. A value of $\lambda=1$ implies
total alignment, i.e. particles
pointing in the direction of streamlines. Finally, the last term on the
right of equation \eqref{Q_motion} accounts for the tendency of the activity to misalign
the nematic, driving it away from equilibrium. 

We project the dynamics of ${\bf Q}$, as described by equations \eqref{Stokes},
\eqref{Q_motion}, onto the space of static
solutions found in the previous section. Taking into account all Fourier
modes that would be an exact representation. To illustrate the approach with a tractable example, 
we consider the 6-dimensional space of solutions,
equations \eqref{Q1_ren_n=0,1},\eqref{Q2_ren_n=0,1} corresponding to restricting our analysis to the first two modes only. In a first step, we linearize 
the equations in ${\bf v}$, $\delta Q_1$, and $\delta Q_2$ 
to obtain
\beqa
&& \eta\triangle^2\psi = -2\left[\left(\alpha+\lambda\Lambda^{2}\right)
+\lambda\triangle\right]\partial_{x}\partial_{y}\delta Q_{1}-
 \alpha\left[\partial_{x}^{2}-\partial_{y}^{2}\right]\delta Q_{2} \nonumber \\ 
&& \quad \quad \quad
+\left[(1-\lambda)\partial_{x}^{2}+(1+\lambda)\partial_{y}^{2}\right]
\triangle\delta Q_{2} \label{eq:dyn_psi_nondim-1} \\
&& \partial_t\delta Q_1=\lambda(\partial_{x}\partial_{y}\psi)+
\triangle\delta Q_{1}-\Lambda^{2}\ \delta Q_{1}
\label{eq:dyn_Q1_psi-1} \\
&& \partial_t\delta Q_2  = 2\left[(\lambda+1)\partial_{y}^{2}\psi +
  (1-\lambda)\partial_{x}^{2}\psi\right]+\triangle\delta Q_2, 
\label{eq:dyn_Q2_psi-1}
\eeqa
writing the velocity  in terms of the stream function
$\psi$ \cite{LL84a} as ${\bf v} = (\partial_y\psi,-\partial_x\psi)$. 

We expand in the small parameters $\lambda$ and
$\alpha$, since for $\lambda = \alpha = 0$ the equations of motion
reduce to the equilibrium case, with no motion. At each order
$\lambda^n\alpha^m$ in an expansion in the two variables, we can the
derive an equation of motion for the coefficients of the equilibrium
solutions. First, we expand each of the coefficients into a Taylor
series in $\lambda,\alpha$, which results in a corresponding series
for $\delta Q_{1/2}$:
\(
\delta Q_{1/2} = \lambda \delta Q_{1/2}^{(\lambda)} +
\alpha \delta Q_{1/2}^{(\alpha)} + \ldots \; ; \; 
\)
and the stream function $\psi$ can be expanded in the same way.
%
As boundary conditions we impose that $\psi^{(\lambda)}$ vanishes at infinity,
and satisfies the no-slip condition
$\psi^{(\lambda)} = \partial_r\psi^{(\lambda)}= 0$
on $r = 1$ \cite{HB83}, corresponding to the microscopic defect core. 
This condition fixes a frame of reference in which the imposed defect
is at rest. 
%
We perform the expansion to order $\lambda^2$
and $\lambda\alpha$ yielding equations of motion for the parameters, 
$E_{1/2}(t),D_{1/2}(t),\zeta_{1/2}(t)$,
\beqa
\label{ODE1}
&& \dot{E}_1(t)=\left(\bar\lambda^2+\alpha\bar\lambda\right)
\frac{1-E_{1}^{(0)}}{4\eta}, \quad \dot{E}_2 = 0, \\
\label{ODE2}
&& \dot{D}_1(t)=-\frac{D_1^{(0)}}{4\eta}\left[\alpha\bar\lambda-
\bar\lambda^{2}\sec\left(2\zeta_{1}^{(0)}\right)\right] \\
\label{ODE3}
&& \dot{\zeta}_1(t)=\frac{\bar\lambda^{2}}{4\eta}
\tan\left(2\zeta_{1}^{(0)}\right) \\
&& \dot{D}_2(t)  = \frac{D_{1}^{(0)}}{4\eta}
\left[-2\bar \lambda\cos\left(\zeta_{1}^{(0)}+\zeta_{2}^{(0)}\right)\right.
\nonumber\\
\label{ODE4}
&& \left. +\alpha\sec2\zeta_{2}^{(0)}\sin\left(\zeta_{1}^{(0)}-
\zeta_{2}^{(0)}\right)\right]  \\
&& \dot{\zeta}_2(t) = \frac{}{2\eta}\frac{D_{1}^{(0)}}{D_{2}^{(0)}}
\left[\bar\lambda\sin\left(\zeta_{1}^{(0)}+\zeta_{2}^{(0)}\right)\right.
\nonumber \\
\label{ODE5}
&& \left.-\alpha\sec2\zeta_{2}^{(0)}\cos\left(\zeta_{1}^{(0)}-
\zeta_{2}^{(0)}\right)\right],
\eeqa
%
whose time-evolution determines the motion of
defects, to be described below. $\bar\lambda=\Lambda\lambda$ is the rescaled inverse length scale emerging from the interplay of alignment and nematic elasticity.
To find the trajectory of defects, one
needs to find the position of their cores by finding the regions where nematic order vanishes by solving for $Q_1 = Q_2 = 0$ at each time
step. 

\begin{figure*}
\includegraphics[scale=0.3]{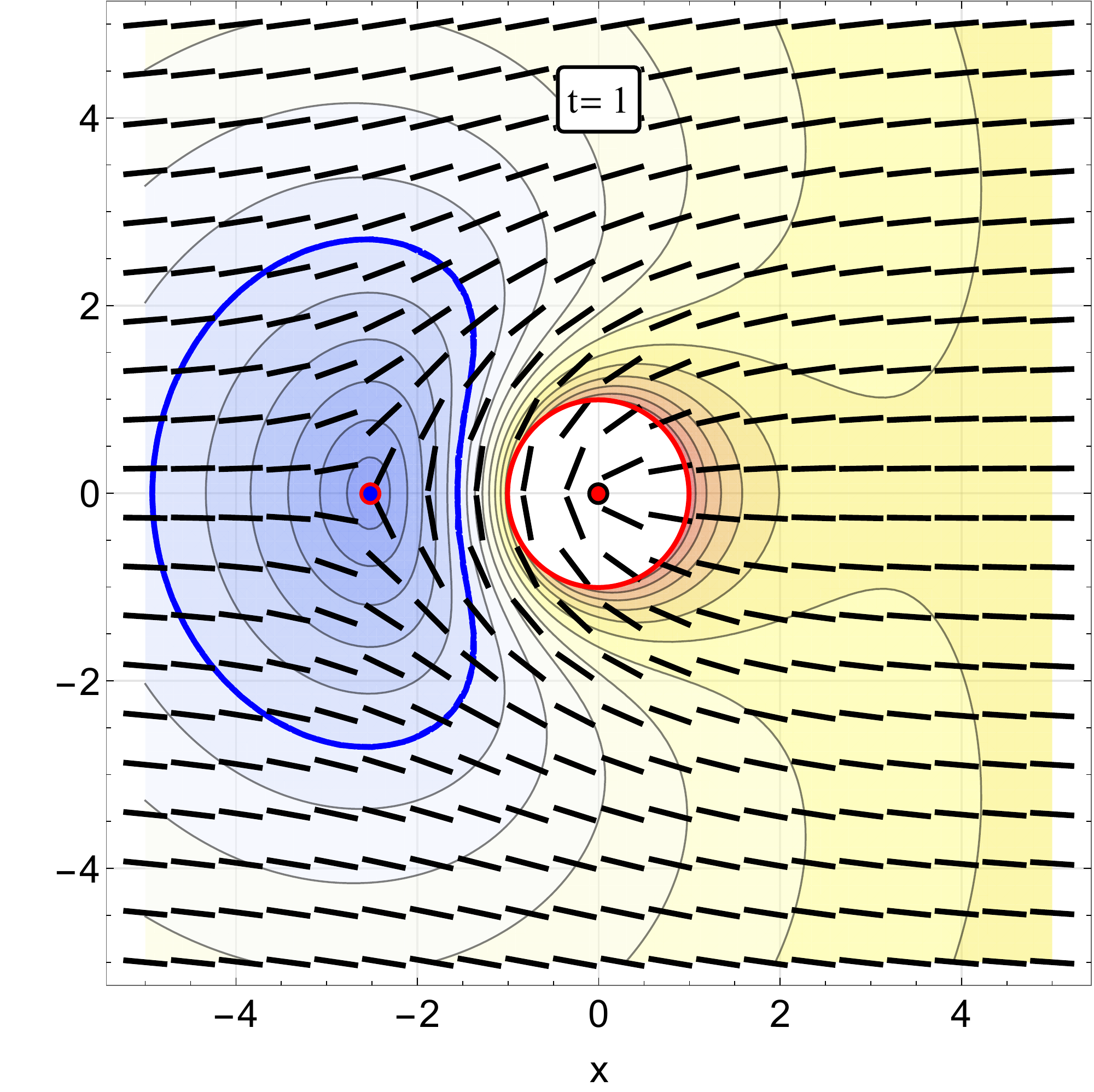}\includegraphics[scale=0.3]{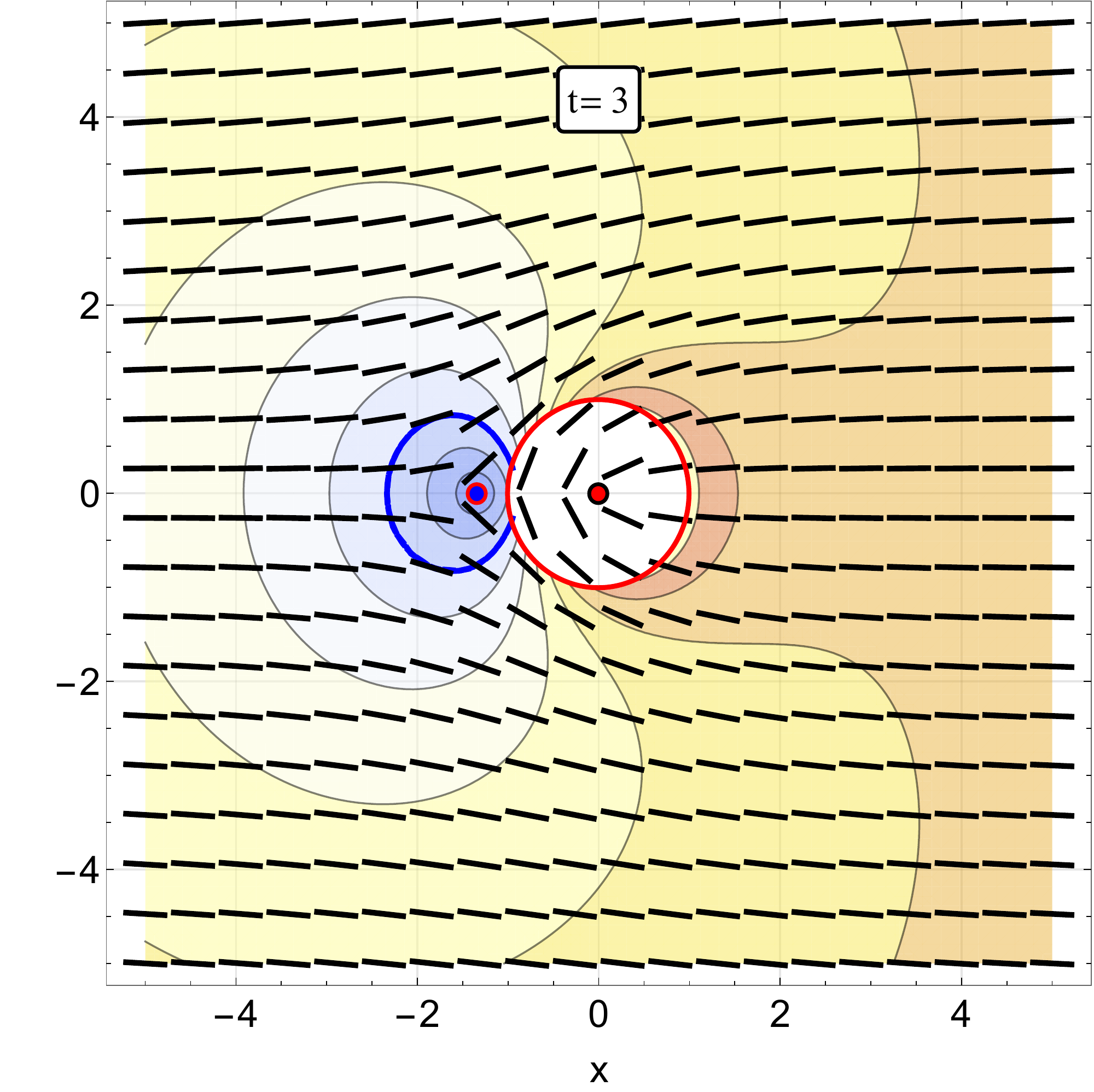}\includegraphics[scale=0.3]{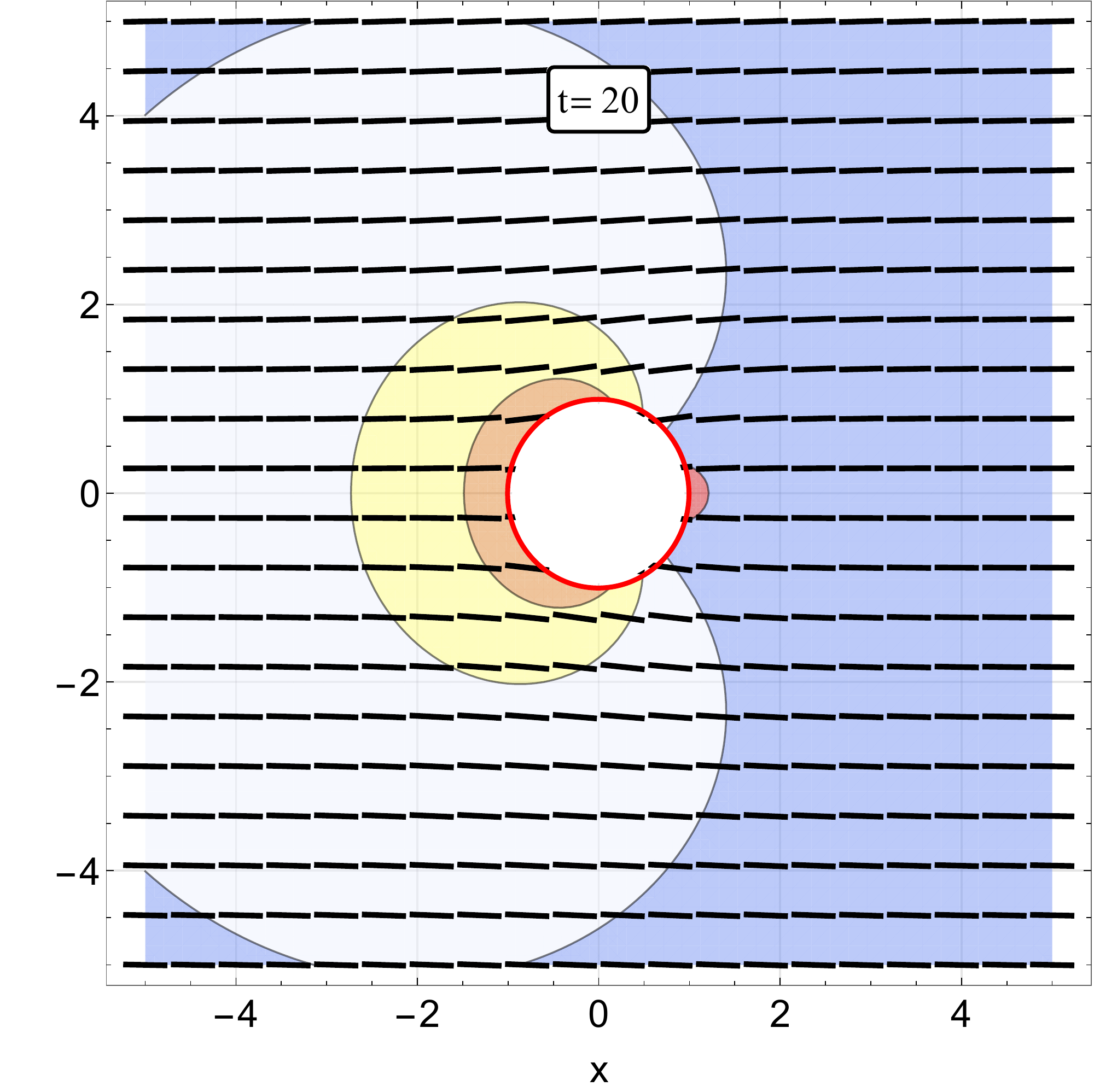}
\caption{Dynamics of a passive nematic, $\alpha=0$. The three panels
show the director and degree of order for $\lambda=0.1$, during the
gradual annihilation of the two defects, that relax onto a state with
uniform director $\mathbf{n}=\mathbf{e}_{x}.$ 
I$(D_1,D_2E_1,E_2,\zeta_1,\zeta_2)^{(0)}=
(0.05,0.05,-0.5,0.1,0,0)$ and $\Lambda = 10^{-3}$.
\label{fig:annihilation}}
\end{figure*}

\begin{figure*}
\includegraphics[scale=0.5]{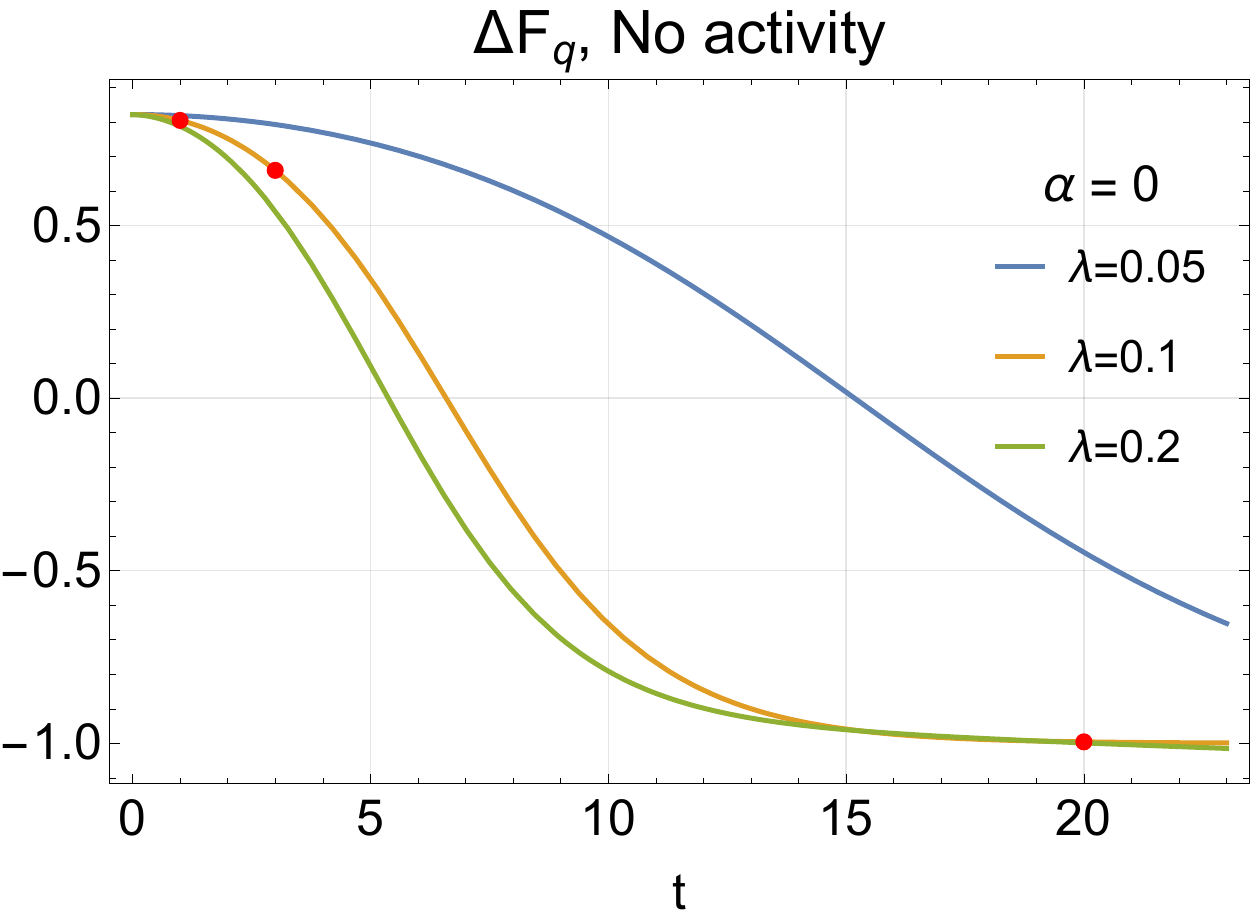}
\caption{The evolution of the Landau-deGennes free energy function as a function of time is plotted for different values of $\lambda.$
The red points correspond to the three profiles plotted above in Figure \ref{fig:annihilation}.
\label{fig:annihilation_ldg}}
\end{figure*}

\subsection{Passive dynamics}
We begin with the dynamics in the absence of activity, $\alpha=0$,
an example of which is shown in Fig.~\ref{fig:annihilation}. The initial
condition is chosen that a pair of 1/2 and -1/2 defects is well separated.
If only alignment effects are present, which are described by terms
proportional to $\lambda$, the systems relaxes to a uniform state.
As seen in Fig.~\ref{fig:annihilation}, the two defects come closer,
until they annihilate (the distance between them becomes smaller than the core size) and the orientation becomes uniform. 

 In Fig. ~\ref{fig:annihilation_ldg}, we have also plotted the Landau-deGennes
free energy, equation \eqref{Free_energy_Q} as a function of time, which is seen to
decrease monotonically. 
As a uniform state is reached, the Landau-deGennes free energy approaches
a constant value. The relaxation toward the uniform value becomes slower as the alignment parameter decreases. 

\begin{figure}
\includegraphics[width=0.8\columnwidth]{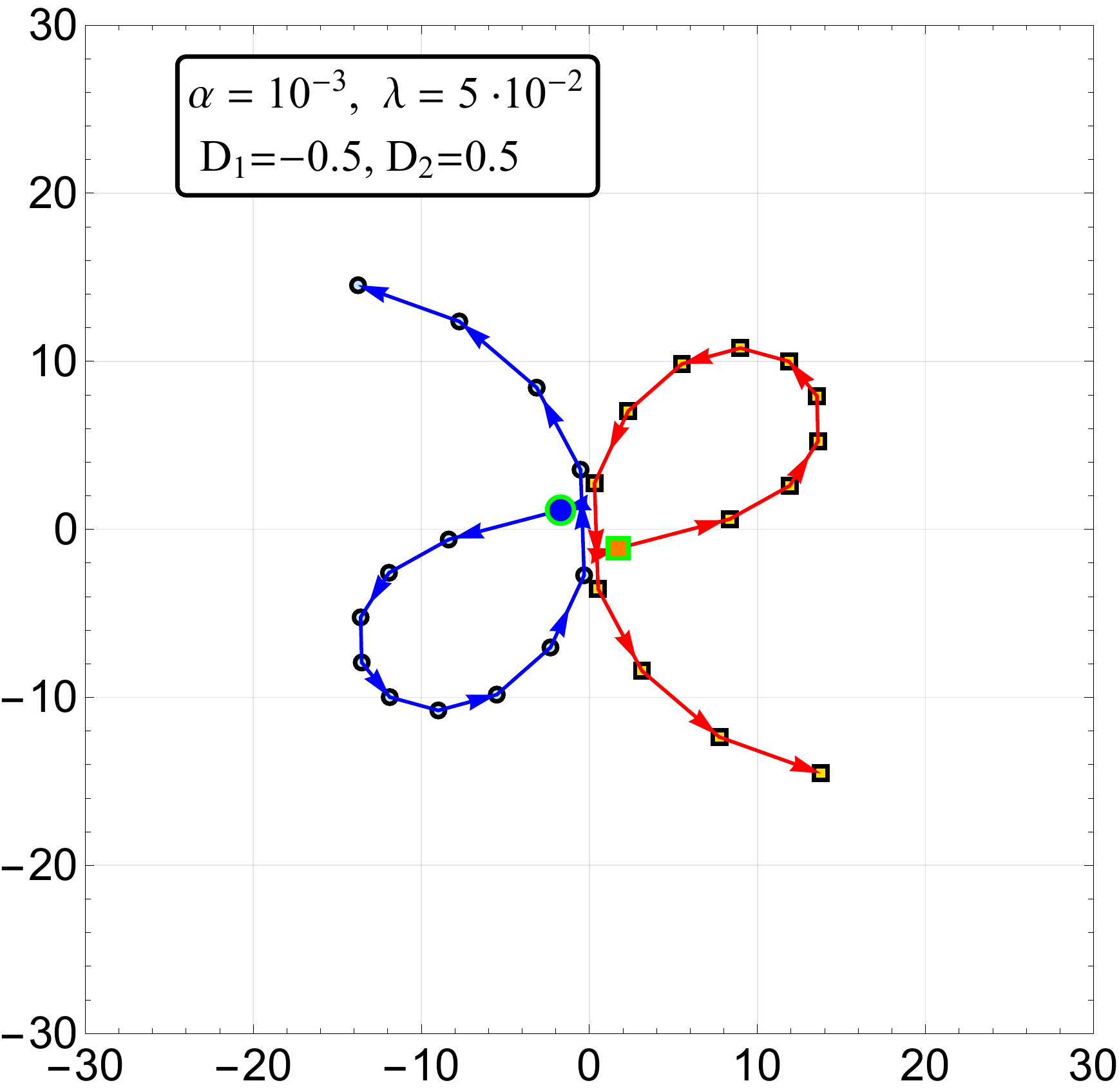}
\caption{Trajectory of a pair of defects (in the frame of reference
of their center of mass) in the presence of low activity,
$\alpha=-10^{-3}$, $\lambda=10^{-2}$. Circles represent -1/2 and squares
+1/2 defects. As the initial configuration (green, larger markers) evolves,
the disclinations trace a spiral, annihilating, then creating a new pair and  growing further apart
several times.
\label{pair_trajectory}}
\end{figure}

\subsection{Active Dynamics}

Next we consider the case where both $\lambda$ and $\alpha$ are
nonzero. Finite activity ($\alpha\ne 0$) pumps energy into the system,
so we expect
defects to be created. On the other hand there is competition with
the alignment terms, which cause defects to annihilate. This is indeed
seen in Fig.~\ref{pair_trajectory}, where the two defects are seen
with their center of mass at the origin. The initial condition is marked
by green squares. At first the two defects move away from one another,
but eventually they turn and come closer to one another, and annihilate,
as their distance becomes smaller than the core size. However, a new pair
is created immediately, starts to move apart, and the process repeats
itself. 
This corresponds very well to what is observed by
\cite{Sanchez2012,idecamp_orientational_2015,Wu2017}, where typically annihilation is followed
immediately by creation of a new pair. 
This dynamics are characterised by a rotational component (governed by $\zeta_{1/2}$) and a radial one (governed by the parameters $E_1$ and $D_1$); as they approach one another or move apart, pair of defects trace spiral-like trajectories (shown in Fig.\ref{pair_trajectory}). 
\begin{figure}
\includegraphics[width=1\columnwidth]{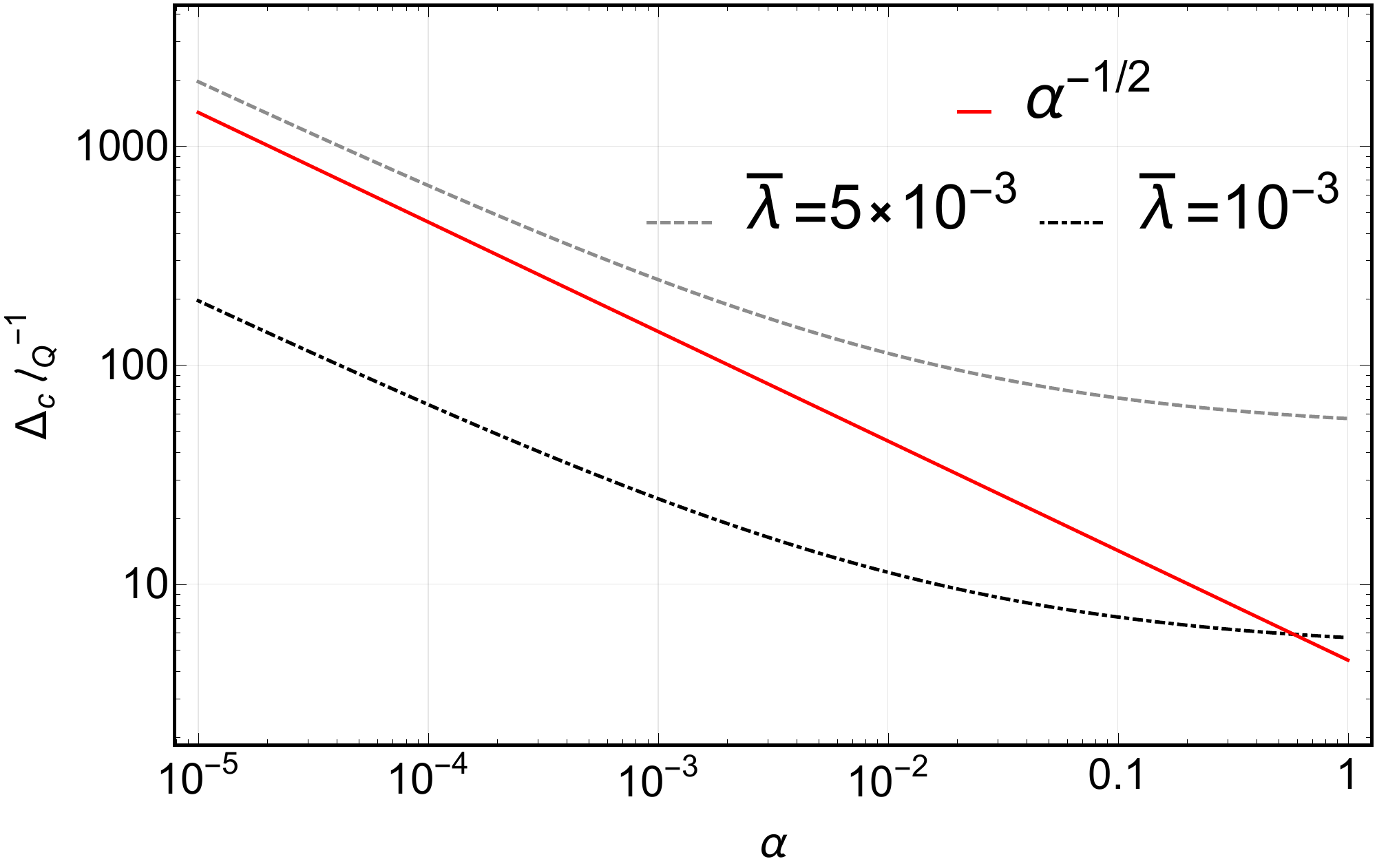}
\caption{The average separation between defects plotted as a function of activity.
 The value of $\lambda$ is set to 0.1.
 \label{average_separation}}
\end{figure}
%
The creation and annihilation of defects will eventually lead to a steady-state 
density of defects when the creation and annihilation balance out. 
This implies an average distance between the defect cores, $\Delta$  
 (the inverse of which determines the density of
defects). 
We estimate this distance by considering a pair of defects at varying initial distances from each other and numerically finding the critical initial distance for which the they neither approach nor repel each other. 
At small values of $\alpha$, we predict a scaling law
$\Delta \propto \alpha^{-1/2}$ which has been observed previously numerically in \cite{hemingway_correlation_2016}.


It is possible to understand the scaling $\alpha^{-1/2}$ by examining
the equations for the dynamics of $Q_{1},$ which is the field that
governs the distance between the two defects in a pair. Keeping 
the terms with lowest order gradients, the equations read
\begin{eqnarray}
\eta\nabla^{4}\psi & = & -2\left(\alpha+\lambda\Lambda^{2}+\lambda\nabla^{2}\right)\partial_{x}\partial_{y}\delta Q_{1}-\alpha\left(\partial_{x}^{2}-\partial_{y}^{2}\right)\delta Q_{2}\nonumber \label{eq:dyn_psi_nondim-1-1}
\end{eqnarray}
\begin{equation}
{\normalcolor \partial_{t}\delta Q_{1}}=\lambda(\partial_{x}\partial_{y}\psi)+\nabla^{2}\delta Q_{1}-\Lambda^{2}\ \delta Q_{1}. \nonumber \label{eq:dyn_Q1_psi-1-1}
\end{equation}
It is evident that the balance between $-\alpha$, $\lambda\nabla^{2}$
and $\lambda\Lambda^{2}$ in the first equation sets a length scale $\Delta$, defined by
\begin{equation}
\alpha\sim\lambda\left(\frac{1}{\Delta^{2}}+\Lambda^{2}\right)\sim\lambda\left(\frac{1}{\Delta^{2}}+\frac{1}{\ell_{Q}^{2}}\right).\label{eq:balance_lengthscales}
\end{equation}
In the regime where $a=1\ll\Delta\ll\ell_{Q},$ this translates into
the scaling law
\[
\Delta\sim\alpha^{-1/2}\sim\ell_{\alpha},
\]
which accounts for the behaviour observed in Fig. \ref{average_separation} 
for small $\alpha.$ As the active parameter increases, the relative
distance between defects becomes comparable to $a=1$, this scaling approximation
breaks down (as the distance $\Delta$ plateaus towards $\Delta=a=1).$

\section{Discussion}
We have formulated a theory for the evolution of the macroscopic structure of a (possibly active) nematic
liquid crystal built on a first-principles description of its singularities (topological defects). The dynamics
are described principally by the motion of the defects contained in a
particular state; however, our equations are for the coefficients of
an expansion in modes, and the position of the defects follow as a
secondary quantity. 

Finally, our model allows for a theoretical prediction of the defect areal density that characterises the chaotic states observed in \cite{Sanchez2012,idecamp_orientational_2015}. Our result shows a scaling that agrees with that derived by \cite{hemingway_correlation_2016} via numerical simulations of the same equations. 

In view of experiments and simulations it would be
interesting to describe states with many defects. Although in principle, by adding more modes in our expansion, we can describe states with an
arbitrary number of defects, it remains to be seen if this will be 
practical. An alternative might be to construct superpositions of states 
made up of {\em pairs} of equal and oppositely charged defects,
which ensures that these states
can be matched to each other without encountering any singularities in the fields. 

Most interestingly, the methods we have used can easily be generalised to analyse groups of topological defects that can be found in a 
variety of field theories whose dynamics can be described by partial differential equations. Natural examples would be 
vortices in XY-models, polar liquid crystals or Newtonian fluids. Higher charge defects can also be studied simply by specifying the appropriate boundary condition at the imposed defect core. Another interesting direction is the study of  populations of defects where the vector field lives on a topologically non-trivial manifold such as a sphere~\cite{Keber2014}.

\begin{acknowledgments}
We are grateful to Y. Ibrahim and V. Slastikov for helpful discussions.
TBL acknowledges support of BrisSynBio, a BBSRC/EPSRC Advanced Synthetic Biology
 Research Centre (grant number BB/L01386X/1).
\end{acknowledgments}

\bibliography{Bibliography,library_sep2017}

\appendix
\section{Appendix: Generating more defects}

\begin{figure}
\includegraphics[width=0.8\columnwidth]{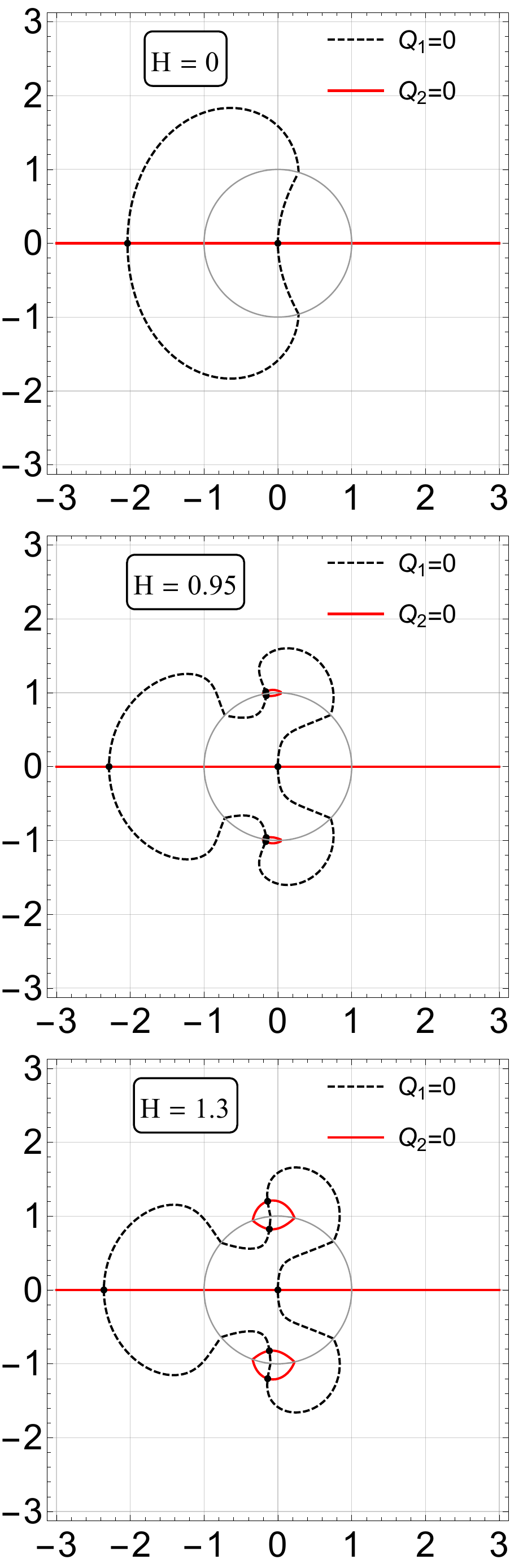}
\caption{Level lines $Q_{1,2}=0$ (dashed black, solid red respectively) for a solution
with three modes. The gray line  indicates the core boundary
$r=1$. The intersection points (black dots)  represent the positions
of the topological defects. As the magnitude of the third mode $H$
increases, the level lines change shape and new pairs of defects appear.
In the central panel two extra pairs are produced at the interface $r=1$;
on the right we see that by varying $H$ the positions of different pairs
and of the single disclinations within pairs changes.
Here $G_{1,2} = 0.1, E_{1,2}=0,\ D_{1,2}=0.9,\ \zeta_{1,2}=0$ and $H_{1,2}=H$ in equations \eqref{Q1_ren_n=0,3}, \eqref{Q2_ren_n=0,3}.
\label{production}}
\end{figure}

While the discussion in the manuscript has mainly considered a single non-zero, i.e. $n=1$ mode only, the analysis can be extended to higher modes.
As an example, in Fig.~\ref{production} we
show the evolution of solutions that have three allowed modes $n=1,2$ and 3: 
\beqa
\label{Q1_ren_n=0,3}
&& \delta Q_1 = (E_1-1)\frac{K_{0}\left(\Lambda r\right)}
{K_{0}\left(\Lambda \right)} + D_1 \frac{K_1\left(\Lambda r\right)}
{K_1\left(\Lambda\right)}\cos(\phi + \zeta_1)  \\ &&
+  G_1 \frac{K_2\left(\Lambda r\right)}
{K_2\left(\Lambda\right)}\cos(2 \phi + \zeta_1) + H_1 \frac{K_3\left(\Lambda r\right)}
{K_3\left(\Lambda\right)}\cos(3\phi + \zeta_1)\; ,   \nonumber \\
\label{Q2_ren_n=0,3}
&& \delta Q_2 = 
E_2 f_2(r,\phi) + D_2\frac{\sin(\phi + \zeta_2)}{r}  \nonumber \\ && +  G_2\frac{\sin(2\phi + \zeta_2)}{r^2}  + H_2\frac{\sin(3\phi + \zeta_2)}{r^3}  \; .
\eeqa

Starting with two defects, (modes $n=2,3$ zero)
 it shows the bifurcations leading to the production of two more pairs of defects. Our analysis indicates that 
$n$ defect pairs can be created with $n$ modes. 

%

\end{document}